\documentclass[twocolumn,aps,pra]{revtex4-2}
\usepackage{graphicx}
\usepackage{dcolumn}
\usepackage{bm}
\usepackage{siunitx}
\usepackage{hyperref}
\usepackage{color}
\usepackage{ulem}
\usepackage{amsthm,amsmath}

\usepackage[colorinlistoftodos]{todonotes}
\presetkeys{todonotes}{inline,backgroundcolor=yellow}{}

\begin{document}

\title{Towards CP Violation Studies on Superheavy Molecules: Theoretical and Experimental Perspective}

\author{R. Mitra$^{a,b}$, V. S. Prasannaa$^c$, R. F. Garcia Ruiz$^d$, T. K. Sato$^e$, M. Abe$^f$, Y. Sakemi$^g$, B. P. Das$^{c}{^,}^{h}$ and B. K. Sahoo$^a$}

\affiliation{$^a$Atomic, Molecular and Optical Physics Division, Physical Research Laboratory, Navrangpura, Ahmedabad 380009, India}
\affiliation{$^b$Indian Institute of Technology Gandhinagar, Palaj, Gandhinagar 382355, India}
\affiliation{$^c$Centre for Quantum Engineering, Research and Education, TCG CREST, Salt Lake, Kolkata 700091, India}
\affiliation{$^d$Massachusetts Institute of Technology, Cambridge, Massachusetts 02139, USA}
\affiliation{$^e$Advanced Science Research Center, Japan Atomic Energy Agency, Tokai, Ibaraki 319-1195, Japan}
\affiliation{$^f$Department of Chemistry, Graduate School of Science, Hiroshima University, 1-3-2, Kagamiyama, Higashi-Hiroshima City, Hiroshima, Japan 739-8511}
\affiliation{$^g$Center for Nuclear Study, The University of Tokyo, Hongo, Bunkyo, Tokyo 113-0033, Japan}
\affiliation{$^h$Department of Physics, Tokyo Institute of Technology,
2-12-1-H86 Ookayama, Meguro-ku, Tokyo 152-8550, Japan}

\date{\today}

\begin{abstract}
Molecules containing superheavy atoms can be artificially created to serve as sensitive probes for study of symmetry-violating phenomena. Here, we provide a detailed theoretical study for diatomic molecules containing the superheavy lawrencium nuclei. The sensitivity to time-reversal violating properties was studied for different neutral and ionic molecules. The effective electric fields in these systems were found to be about 3-4 times larger than other known molecules on which electron electric dipole moment experiments are being performed. Similarly, these superheavy molecules exhibit an enhancement of more than 5 times for parity- and time-reversal-violating scalar-pseudoscalar nucleus-electron interactions. We also briefly comment on some experimental aspects by discussing the production of these systems. 
\end{abstract}
\maketitle 

\section{Introduction}

Recent advances in studies on heavy molecular systems are enabling new opportunities in fundamental physics~\cite{ACME, ThO, MandB1, BaF1, BaF2, Kara, RaF, HgF, HgA,YbOH,RaOH,HgOH,HgF''}. Symmetry-violating properties of nuclei and fundamental particles can be highly enhanced in certain molecules. Precision experiments with ThO molecules, for example, have provided the most stringent bound to electric dipole moment (EDM) of an electron \cite{ACME}, constraining the existence of new physics at the TeV scale \cite{mssm1}. The search for a permanent EDM of an elementary particle has attracted a great deal of experimental and theoretical attention as it is highly sensitive to new sources of time-reversal (T) violation (equivalent to CP violation) ~\cite{CP}, knowledge of which can be useful in explaining the matter-antimatter asymmetry in the Universe~\cite{masm1,masm2}. 

In molecules, the sensitivity to symmetry-violating phenomena scales with the mass and charge of the nuclei \cite{Schiff}. Hence, molecules containing super-heavy nuclei could serve as particularly sensitive laboratories. Future studies of these systems can open up new opportunities in fundamental physics and chemistry. There have been recent advances in the search for the electron EDM using paramagnetic molecules, which would be a signature of CP violation beyond the standard model of elementary particles. Moreover, spectroscopic results from studies of superheavy elements have recently been reported, and in view of these developments, it is desirable to investigate  EDMs using molecules with superheavy atoms. However, our experimental knowledge of such rare molecules is in its infancy, and theoretical developments are critical to motivate and guide the experimental progress \cite{Ga20,Ud21}. 

Within a molecule, the magnitude of the intrinsic electric field that an electron with an EDM experiences, due to the other electrons and nuclei, can be viewed as an effective electric field ($\mathcal{E}_{\mathrm{eff}}$) \cite{Schiff}.
Because of their large atomic number, super-heavy radioactive elements are expected to exhibit a substantial enhancement in their $\mathcal{E}_{\mathrm{eff}}$, since one expects relativistic and electron correlation effects to be prominent. Indeed, effective electric fields of CnH~\cite{CnHIsaev, CnH'}, CnF~\cite{MF}, LrO, NoF, RfN, E120F, and E121O~\cite{Radicals} have been calculated, and asserts the above statement. In this work, we focus our studies on diatomic molecules containing Lr atoms, i.e. LrO, LrF$^+$, and LrH$^+$. Because of short half-lives and the low production rates, superheavy elements, including Lr, need to be handled on a single-atom scale. There are fourteen different isotopes of Lr, of which we propose to employ $^{256}$Lr (half-life $\approx$ 27 s) that has been used in several studies of atomic properties. A purified single ion-beam of $^{256}$Lr, synthesized in the $^{249}$Cf ($^{11}$B, $4n$) reaction, has been successfully produced by using the ISOL (Isotope Separator On-Line) system at the Tandem accelerator facility of Japan Atomic Energy Agency (JAEA)~\cite{Sato2}. We explore a possible alternative scheme to produce Lr atoms. Our work also aims to extend the studies of molecular ions, which could enable precision measurements with just a single molecular ion \cite{Ca17,Yu21,Fa21}. 

The manuscript is organized as follows:  Sec. \ref{theo} presents the general theory of molecular EDMs due to electron EDM and scalar-pseudoscalar (S-PS) nucleus-electron interactions. The many-body method employed in the present work is explained in Sec. \ref{method}. Our theoretical results and discussions about the sensitivity of these molecules to symmetry-violating properties are presented in  Sec. \ref{results}. 

\section{Theory} \label{theo}
Theoretical studies for the sensitivity to symmetry-violating properties have been performed for the diatomic molecules LrO, LrF$^+$, and LrH$^+$. We employ a 4-component relativistic coupled-cluster (RCC) method for this purpose. The fully relativistic aspect becomes especially relevant, as theoretical calculations have predicted that the Lr atom would have a configuration different from that expected by a non-relativistic treatment, [Rn]$5f^{14}6d^17s^2$, due to strong relativistic effects~\cite{Naito}. The [Rn]$5f^{14}7s^27p_{1/2}$ configuration would be most probable for the Lr atom according to the measurement of the first ionization potential measurement~\cite{Sato2}. We employ the configuration for calculations in present work. Since LrO, LrF$^+$ and LrH$^+$ have one unpaired electron each, these molecules are sensitive to both the electron EDM and nucleus-electron S-PS interactions~\cite{Bouchiat}. Thus, these T-odd sources can induce an energy shift in the molecular levels given by 
\begin{eqnarray}
\Delta E \simeq -d_e \mathcal{E}_{\mathrm{eff}}+ k_s W_s ,\label{Eqq}
\end{eqnarray}
where the first term on the right hand side of Eq. (\ref{Eqq}) describes the contribution from electron EDM and the second term shows that from the S-PS interaction. In the above equation, $k_s$ is the coupling parameter for S-PS interaction, with the accompanying quantity $W_s$ being the analogue of $\mathcal{E}_{\mathrm{eff}}$, and like the effective electric field, it too can only be calculated using many-body theory. 

The interaction Hamiltonian due to electron EDM in a molecule is given by \cite{24}
\begin{eqnarray}
H_{EDM} &=& -2 icd_e \sum_{j=1}^{N_e}\beta\gamma_5p_j^2 \nonumber \\
  &=& \sum_{j=1}^{N_e} h_{EDM,j} , 
\end{eqnarray}
where $c$ is the speed of light, $\beta$ is a Dirac matrix, $\gamma_5$ is the product of Dirac matrices, $p_j$ is the momentum operator corresponding to the $j^{th}$ electron, and $N_e$ is the total number of electrons in the system. Assuming that the shift in energy is only due to the electron EDM, the expression for $\mathcal{E}_{\mathrm{eff}}$ can be obtained as  
\begin{eqnarray}
\mathcal{E}_{\mathrm{eff}}=-\frac{1}{d_e}\frac{\langle\Psi\arrowvert H_{EDM}\arrowvert\Psi\rangle}{\langle\Psi\arrowvert\Psi\rangle}\label{2},
\end{eqnarray}
where $\arrowvert\Psi\rangle$ is the ground state wave function of the molecule. Similarly, the S-PS interaction Hamiltonian manifested at molecular level is given by~\cite{Lindroth,24,kozlov} $H_{S-PS}= \sum_{A=1}^{N_{nuc}}H_{S-PS,A}$, with
\begin{eqnarray}
H_{S-PS,A} = i\frac{G_F}{\sqrt {2}}k_{s,A}Z_A\sum_{j=1}^{N_e}\beta\gamma_5 \rho_A(r_{A_j}) ,
\end{eqnarray}
where $A$ denotes the $A^{th}$ nucleus, $Z_A$ is the nuclear charge of the $A^{th}$ nucleus, $\rho_A$ the nuclear charge density,  $G_F=2.219 \times 10^{-14}$ (in atomic units (a.u.)) is the Fermi coupling constant, $N_{nuc}$ is the total number of nuclei in the molecule, and $k_{s,A}$ is the S-PS coupling coefficient of the corresponding nucleus, $A$. We have accounted here contribution only from the heaviest atom (since the contribution from the heavier of the two atoms dominates) and denote $k_{s,A}$ as $k_s$ for brevity. The quantity of theoretical interest in the S-PS interaction, $W_s$, are given by 
\begin{eqnarray}
W_s=\frac{1}{k_s}\frac{\langle\Psi\arrowvert H_{S-PS}\arrowvert\Psi\rangle}{\langle\Psi\arrowvert\Psi\rangle}\label{3} .
\end{eqnarray} 
We consider the Gaussian nuclear charge distribution to define nuclear charge density, given by $\rho_A(r) = \left(\frac{\eta_A}{\pi}\right)^{\frac{3}{2}} e^{-\eta_A r^2}$, with $\eta_A=\frac{3}{2}(R_{A}^{rms})^{-2}$, where $R_A^{rms}$ is the root mean squared nuclear charge radius of the $A^{th}$ nucleus. 

Another property that plays an important role in determining the statistical sensitivity of a molecule in an EDM experiment is the molecular permanent electric dipole moment (PDM). The PDM can be evaluated as the expectation value of the electric dipole operator, given as
\begin{eqnarray}
\mu &=&\sum_{A=1}^{N_{nuc}}Z_A R_A-\sum_{i=1}^{N_e}r_i \nonumber \\
&=& \mu_{nuc} - \mu_e,
\end{eqnarray}
where $R_A$ and $r_i$ are respectively the position of the $A^{th}$ nucleus and $i^{th}$ electron with respect to the origin. Since we employ the Born-Oppenheimer approximation~\cite{BO}, where the nuclei are clamped, $R_A$ are constants, and are determined from the minimum of the potential energy surface for a given molecule. In the expression, $\mu_{nuc}$ and $\mu_e$ are the nuclear and electronic contributions to PDM, respectively. Since the investigated molecules are diatomic systems, each of the systems have only one equilibrium bond length ($R_e$). We have set the Lr atom as the origin to define $R_e$. Therefore, the PDMs of the charged systems are specified with Lr as the origin. We add here that it is straightforward to obtain the PDM at any other origin, including setting it at the centre of mass of the molecular system, as the property simply scales with change in origin. 

\section{Method Of Calculation}\label{method}

The molecular parameters were calculated by solving the many-body wave function of the molecular system within a relativistic framework. To account for the relativistic  and electron correlation effects rigorously, we employ here the RCC theory, expressing the wave function as \cite{Bishop}
\begin{eqnarray}
\arrowvert\Psi\rangle=e^T\arrowvert \Phi_0 \rangle,
\end{eqnarray}
where $\arrowvert\Phi_0\rangle$ is the Dirac-Hartree-Fock (DF) wave function, and $T$ is the cluster operator. The latter is responsible for generating particle-hole excitations which arise out of the DF state due to the residual Coulomb interaction that is ignored in the evaluation of the DF wave function. $T$ is expressed as
\begin{eqnarray}
T=T_1+T_2+T_3+\cdots+{T_N}_e,
\end{eqnarray}
where the operator $T_i$ accounts for all possible $i$ particle- $i$ hole excitations. Due to the prohibitively expensive cost associated with including all the excitations, we adopt the common practice where one restricts $T$ to include only singles and doubles excitations (RCCSD method), in which the excitation operators are defined using the second quantization operators as
\begin{eqnarray}
T_1 &=& \sum_{i,a} t_i^a a_a^\dagger a_i,
\end{eqnarray}
and
\begin{eqnarray}
T_2 &=&\frac{1}{4}\sum_{i,j,a,b}t_{ij}^{ab}a_a^\dagger a_b^\dagger a_ia_j,
\end{eqnarray}
where the indices $i$, $j$, $k$, $\cdots$ etc. denote the occupied orbitals, and $a,b,c,\cdots$ identify the virtual orbitals, $t_i^a$ is the amplitude for a single excitation from $i^{th}$ occupied orbital to $a^{th}$ virtual orbital, whereas, $t_{ij} ^{ab}$ is the amplitude of the doubles excitation from the occupied $i$ and $j$ orbitals to the virtual $a$ and $b$ orbitals, respectively.

Once the ground state wave function of a molecule is obtained, we calculate the properties of interest by using the expectation value approach. The expectation value of an operator $O$ in the RCC method is given by 
\begin{eqnarray}
\langle O\rangle&=&\frac{\langle\Phi_0\arrowvert e^{T^\dagger} O e^T\arrowvert \Phi_0\rangle}{\langle \Phi_0\arrowvert e^{T^{\dagger}}e^T \arrowvert\Phi_0\rangle}\nonumber\\
&=&\langle\Phi_0\arrowvert e^{T^\dagger}O e^T\arrowvert \Phi_0\rangle_l,
\end{eqnarray}
where the subscript $l$ means that only the linked terms contribute. In the RCCSD method, since $T\simeq T_1 +T_2$, we have
\begin{eqnarray}\label{expecvalue}
\langle O \rangle &=& \langle \Phi_0 \arrowvert O \arrowvert \Phi_0 \rangle_l + \langle \Phi_0 \arrowvert  (OT_1 + h.c.) + T_1^{\dagger} O T_1 + T_2^{\dagger} O T_2  \nonumber \\ && + \frac{1}{2} (T_1^{\dagger} O T_1^2 +h.c.) + \frac{1}{2}  (T_2^{\dagger} O T_1^2 +h.c.) \nonumber \\
  && + (T_2^{\dagger} O T_1 T_2 +h.c.) + \cdots \arrowvert \Phi_0 \rangle_l ,
\end{eqnarray}
where h.c. stands for the hermitian conjugate term \cite{Cizek, Bartlett}. It can be noted that $OT_2$ and its h.c. term do not give us  fully contracted terms owing to the one-body form of the operators, which describe the properties that are investigated here, i.e., $O = \mu$, $O=H_{EDM}$ and $O=H_{S-PS}$ for the evaluation of $\mu$, $\mathcal{E}_{\mathrm{eff}}$ and $W_s$ respectively. In the above expression, the first term gives the DF value and the $OT_1$ term includes electron correlations arising through the core-polarization and pair-correlation effects to all orders. Therefore, it contributes dominantly to the electron correlation effects.
 
\begin{figure*}[t]
    \setlength{\tabcolsep}{1mm}
        \begin{tabular}{ccc}
            \includegraphics[height=40mm,width=60mm]{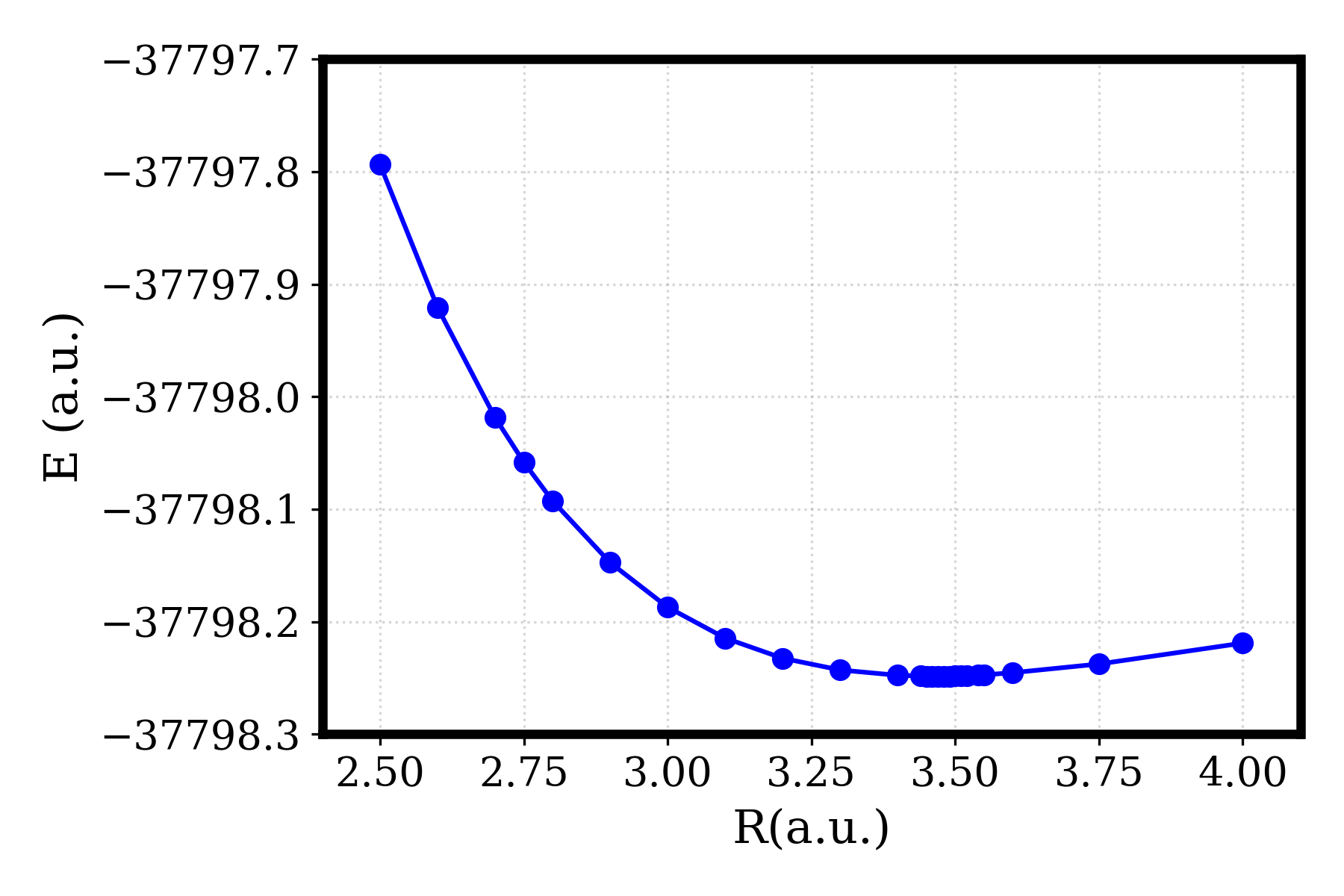} &  \includegraphics[height=40mm,width=60mm]{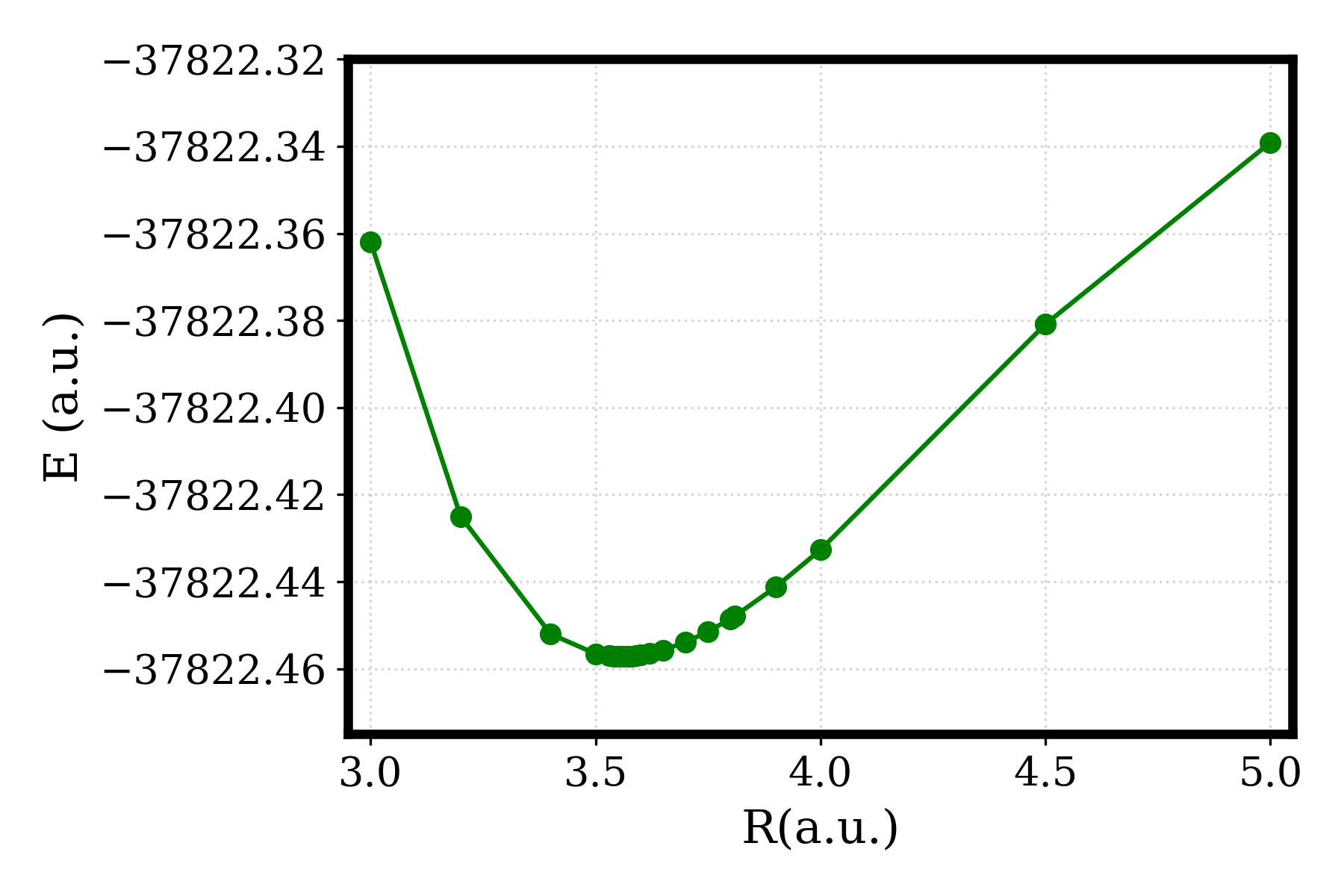}&  \includegraphics[height=40mm,width=60mm]{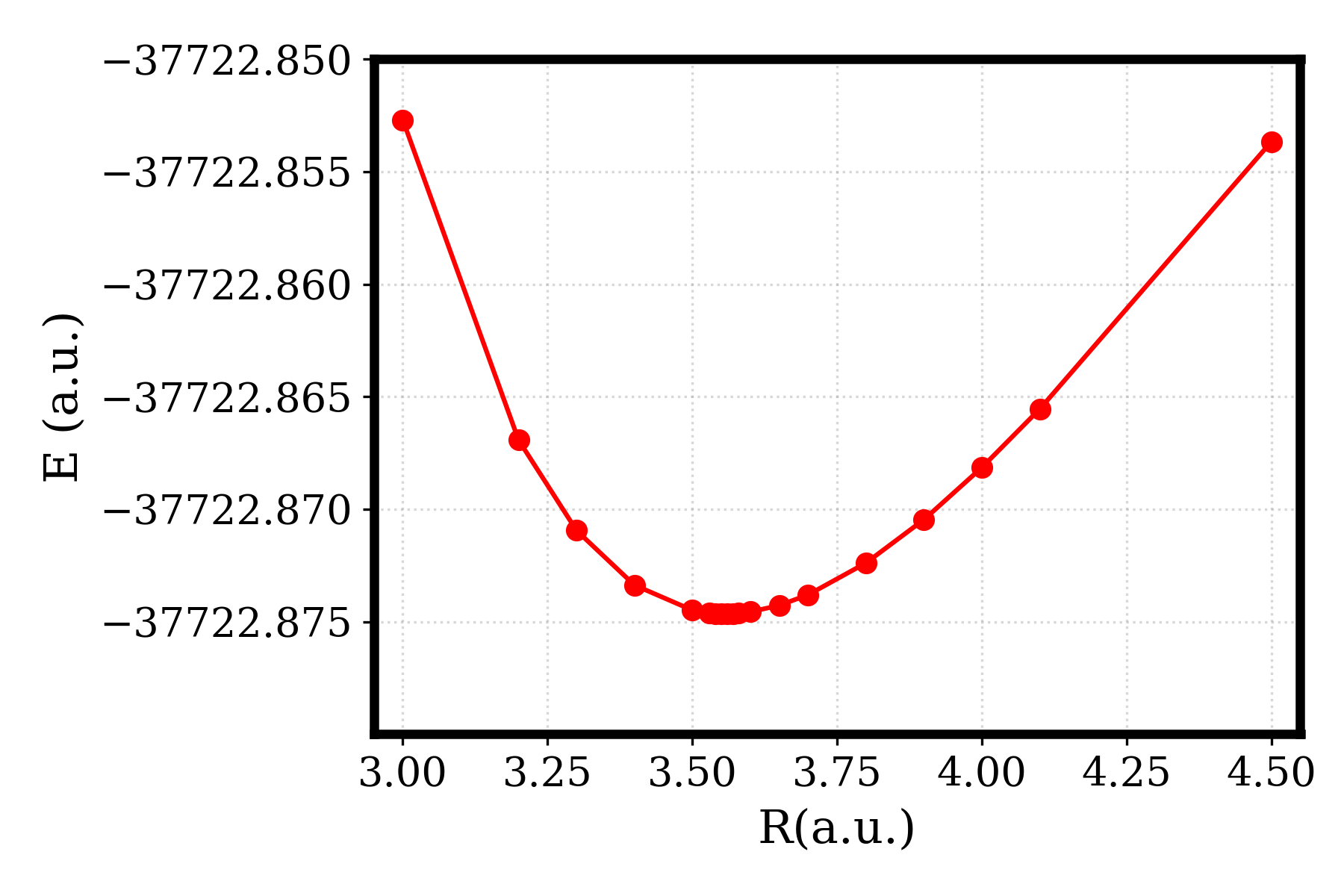}\\
            \textbf{(a)}& \textbf{(b)}& \textbf{(c)}\\
        \end{tabular}
\caption{Potential energy curves (PEC) of neutral LrO (sub-figure (a)), as well as LrF$^+$ (sub-figure (b)), and LrH$^+$ (sub-figure (c)) molecular ions with a range of typical molecular bond-lengths ($R$) using the RCCSD(T) method. The plot shows that LrO can form a stable molecule around the equilibrium bond-length $R_e \simeq 3.46$, while both the ions possess their minima around $R_e \simeq 3.56$. All the values are given in a.u.. }
\label{fig2}
\end{figure*}
 
\section{Results and Discussion}\label{results}
 
At the outset, it is critical to find if the systems of interest can form bound states. The ground electronic state energies at several bond lengths were calculated to construct the potential energy curve (PEC) of each molecule. We then identify the minimum of the PEC, whose corresponding bond-length provides the molecular $R_e$. We carry out this procedure for LrO, LrF$^+$, and LrH$^+$ (Fig. \ref{fig2}), while ensuring that for each molecule, we choose more grid points around the minimum, to pin-point $R_e$ to an accuracy of two decimal places in a.u.. We employ the RCCSD method with the inclusion of partial triple excitations in a perturbational manner, using the Dirac 18 package ~\cite{Dirac18, LVIS}. This approach is referred to as the RCCSD(T) method. We use Dyall's triple-zeta (TZ) v3z~\cite{basis1} basis sets. For lowering the computational requirements while compromising little on accuracy, we have cut off the high-lying virtual orbitals with energies above 2000 a.u.. The value of $R_e$ for LrO was found to be 3.46 a.u., while that of LrF$^+$, and LrH$^+$ were found to be 3.56 a.u.. The PECs of each of the molecules exhibit a smooth trend with a clear global minimum (see Figure \ref{fig2}). 

After finding the equilibrium bond-lengths of the investigated molecules, we calculate the other properties of interest. For this purpose, we use Dyall's quadruple-zeta (QZ) v4z~\cite{basis1} basis sets. We employ the UTChem~\cite{Utchem1,Utchem2} package for DF calculations and for atomic orbital to molecular orbital integral transformations, and in tandem, use the Dirac08 package for RCC calculations~\cite{Dirac08}. We finally use our expectation value code to evaluate the values of the properties~\cite{nleccc}. We present the calculated values of $\mathcal{E}_{\mathrm{eff}}$, $W_s$ and $\mu$ of LrO, LrF$^+$, and LrH$^+$ from the DF and RCCSD methods in Table \ref{tab:1}. 
We also compare our results for LrO with the only available values in literature from Refs.~\cite{Radicals,LrOnew}. In Ref.~\cite{Radicals}, the authors have used (37$s$, 34$p$, 14$d$, and 9$f$) uncontracted Gaussian type functions for Lr, and a decontracted atomic natural orbital (ANO) basis set of TZ quality for O. They performed their calculations by employing complex generalised Hartree-Fock (cGHF) as well as complex generalised Kohn-Sham (cGKS) theories. They obtained an $R_e$ of 3.51 a.u. and 3.53 a.u. with cGHF and the cGKS approaches, respectively, which is in reasonably close proximity to our estimated value of 3.46 a.u.. The value of $\mathcal{E}_{\mathrm{eff}}$ using cGHF method came out to be 322.58 GV/cm, while the cGKS method yielded 250.21 GV/cm. Our calculation, using the RCCSD method and with Dyall's QZ bases for Lr: (37$s$, 34$p$, 24$d$, and 14$f$) and O: (18$s$ and 10$p$), gives an $\mathcal{E}_{\mathrm{eff}}$ of 258.92 GV/cm, and is in better agreement with their results from the cGKS method than the cGHF approach. 
In Ref.~\cite{LrOnew}, the authors use analytic first derivatives for X2C (exact 2-component) CCSD and CCSD(T) methods, and employed TZ quality basis sets. Further, they froze several of the occupied orbitals in their computations, and obtain an effective electric field of 263.9 GV/cm with CCSD and 246.5 GV/cm with the CCSD(T) approaches. Although their CCSD results are in reasonable agreement with our CCSD results obtained using all-electron fully relativistic CCSD calculations and with a QZ basis, it could be fortuitous. This is evident from the disagreement in the correlation trends in Ref.~\cite{LrOnew}, where the DF result is greater than that CCSD counterpart. In Table \ref{tab:1}, we have also compared our results with the corresponding values of ThO, where the most accurate EDM measurement is available, and HgF, which possesses the largest $\mathcal{E}_{\mathrm{eff}}$ estimated thus far for non-super-heavy systems. The Lr molecules were found to have values of  $\mathcal{E}_{\mathrm{eff}}$ that are 3-4 times larger than ThO and HgF. Similarly, the values of $W_s$ were found to be about 5 times larger. 

\begin{table}[t]
\caption{Calculated values of $\mathcal{E}_{\mathrm{eff}}$, $W_s$ and $\mu$ for LrO, LrF$^+$, and LrH$^+$ using the RCCSD method, and comparison with literature values wherever they are available. We also compare these values with calculations of the corresponding quantities for ThO and HgF (see the text for further details). }
\begin{tabular}{l c c c c} 
 \hline \hline
 Molecule& $\mathcal{E}_{\mathrm{eff}}$ (GV/cm) & $W_{s}$ (kHZ) & $\mu$ (D) & Reference \\ 
 \hline \\
 LrO  &  258.92& 1032.23 & 4.58 & This work\\
      & 250.21& 938$^a$ &&\cite{Radicals}\\
      &246.5 & & & \cite{LrOnew}\\
 LrF$^+$  &  246.31& 977.37  & 12.29  & This work \\
 LrH$^+$  & 343.38 & 1375.62 & 11.05 & This work \\
 ThO & 87 & 232$^a$  & 4.27  & Ref. \cite{ThO} \\
 HgF & 115.42 & 264.7 & 2.61 & Ref. \cite{HgF}\\  
 \hline \hline
\end{tabular}
\begin{tabular}{l}
$^a$In Refs.~\cite{Radicals} and~\cite{ThO}, the authors give 469 and 116 kHz \\ for $W_s$, respectively, and we have multiplied their values \\ by a factor of two in the table for reasons pertaining to \\ consistency in notation.\\ 
\end{tabular}
\label{tab:1}
\end{table}

\begin{table*}[t]
\caption{Contributions from the individual terms of the expectation value expression in the RCCSD method to $\mathcal{E}_{\mathrm{eff}}$, $W_s$ and $\mu$ for LrO, LrF$^+$, and LrH$^+$. The table shows the trends in different contributions between $\mathcal{E}_{\mathrm{eff}}$ and $W_s$. }
    \centering
    \begin{tabular}{cccccccccc}
\hline\hline\\
Term &\multicolumn{3}{c}{$\mathcal{E}_{\mathrm{eff}}$ (GV/cm)} &\multicolumn{3}{c}{$W_s$ (kHz)}& \multicolumn{3}{c}{$\mu$ (D)}\\
 \cline{2-4} \cline{5-7} \cline{8-10} \\
& LrO & LrF$^+$ &LrH$^+$ & LrO & LrF$^+$ & LrH$^+$ & LrO & LrF$^+$ &LrH$^+$\\
\hline \\
$O$ &  235.16   & 215.20 &  283.59 & 944.28   & 864.94 & 1135.23& $-$898.40 & $-$828.32&$-$919.79\\
$OT_1+$h.c. & 65.02&  35.52&  76.06  & 254.36  & 140.13 & 300.45 &  $-$1.62 & $-$0.56&$-$0.72\\
$T_{1}^{\dagger} OT_1$ & $-26.28$ & $-$2.24 & $-$9.56 & $-$105.67 & $-$9.02 & -38.91 &$-0.59$ & $-$0.09&$-$0.14\\
$T_1^\dagger OT_2+$h.c. & $-8.11$ & $-$2.87 & 0.95  & $-$33.28  & $-$11.82 & $-$32.96  & 0.58 & 0.14 & 0.12 \\
$T_2^\dagger OT_2$ & $-$8.54 & $-$1.70 & $-$8.21 & $-$34.43 & $-$6.86 & 3.73  &  $-$0.48 & $-$0.14 & $-$0.14 \\
Others & 1.67 & 2.4  & $-$0.45  & 6.97 & 6.7  & 1.38 & $-$0.44 &0.08 &0.02 \\
Nuclear term&$-$&$-$&$-$&$-$&$-$&$-$&905.56&841.27&931.73\\
Total&258.92&246.31&343.38&1032.23&977.37&1375.62&4.58&12.29&11.05\\
\hline\hline
\end{tabular}
    \label{termbyterm}
\end{table*}

In Table \ref{termbyterm}, we present contributions to the values of $\mathcal{E}_{\mathrm{eff}}$, $W_s$ and $\mu$ for LrO, LrF$^+$, and LrH$^+$ from different terms of the RCCSD method, given in Eq.~(\ref{expecvalue}). The first term corresponds to the DF value, while other terms represent correlation contributions. As the table shows, $OT_1+$h.c. terms contribute the most to the correlation effects. We note at this point that $OT_1$ primarily contains in it correlation effects arising from interaction of pairs of electrons. $T_1^\dagger OT_1$ also contains effects involving pairs of electrons, but the interplay between the EDM and Coulomb interactions are more complex. The next leading-order contributions arise from $T_{1}^{\dagger} OT_1$. {\it Albeit} the terms related to $T_2$ operator give comparatively small contributions, they are non-negligible. In fact, for LrO, a sizeable amount of the contributions from $OT_1+$h.c. terms are cancelled out by the other terms that are linear in $T$. The row denoted as `Others' show that the non-linear terms are negligible even for the considered super-heavy systems.  Comparisons between the ratios of the magnitudes of $OT_1+$h.c. and DF values of $\mathcal{E}_{\mathrm{eff}}$, $W_s$, and $\mu$ in LrO, which come out to be $\approx 0.28$, $\approx 0.27$ and $\approx 0.23$, respectively, indicate that the electron correlation effects are almost equally important in all these quantities. Similarly, for  $\mathcal{E}_{\mathrm{eff}}$, $W_s$, and $\mu$, we find these ratios as $\approx 0.16$, $\approx 0.16$ and $\approx 0.27$, respectively, for LrF$^+$, and $\approx 0.27$, $\approx 0.17$ and $\approx 0.26$, respectively, for LrH$^+$. These results suggest that the electron correlation trends are almost similar in both molecular ions. 

We now briefly comment on the molecular orbital information in the chosen systems. We expect that the singly occupied molecular orbital (SOMO) electron is localized in the $s$ orbital of Lr from the following reasoning: as Lr has $7s^27p^1$ configuration, while O, H$^+$, and F$^+$ are expected to pull two electrons towards themselves owing to their larger electronegativity, thus leading to Lr $7s^1$. We verified this reasoning explicitly by examining the atomic components of the SOMO for all three systems, and indeed found that the atomic orbitals from Lr provide the dominant contributions. We also carried out population analysis, and found that in LrF$^+$, for example, the SOMO is predominantly made out of the $s$ function of Lr (0.86), followed by its $d$ (0.09) and $p$ functions (0.0388). 

\begin{figure}[t]
\centering
   \includegraphics[width=8.7cm,height=8.1cm]{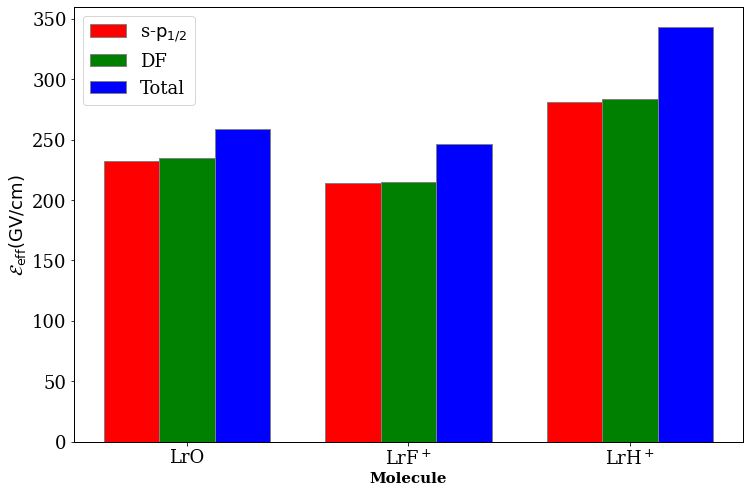}
\caption{Bar plots showing effective electric fields of LrO, LrF$^+$, and LrH$^+$ at three levels: the $s-p_{1/2}$ mixing contributions from Lr at the DF level of theory (as red bars), the DF values themselves (as green bars), and the total values (in blue). All units are in GV/cm. }
\label{fig1}
\end{figure}

As seen in Table \ref{termbyterm}, the major contribution to the properties of interest in this work comes from the DF part. Therefore, we take a closer look at the DF contribution, in order to understand the possible reasons for observing large values of  $\mathcal{E}_{\mathrm{eff}}$ and $W_s$ in the studied superheavy molecules, as compared to other systems. Typically, in these molecules, the heavier atom provides most of the contributions to $\mathcal{E}_{\mathrm{eff}}$ and $W_s$. Due to the short-range and odd-parity nature of the scalar interaction Hamiltonians, the $s_{1/2}$ and $p_{1/2}$ orbitals generally contribute predominantly to $\mathcal{E}_{\mathrm{eff}}$ and $W_s$. It is known that relativistic effects deform the inner core orbitals, $s_{1/2}$ and $p_{1/2}$, strongly in the heavier atomic system. Thus, it is anticipated that these orbitals can strongly influence the $\mathcal{E}_{\mathrm{eff}}$ and $W_s$ values in LrO, LrF$^+$, and LrH$^+$. We explicitly verify this argument by decomposing the DF contribution to $\mathcal{E}_{\mathrm{eff}}$ (we do not repeat the analysis for $W_s$, as we expect for it similar trends as the effective electric field) for all the three systems as
\begin{eqnarray}
\mathcal{E}_{\mathrm{eff}}^{DF} &=& \frac{1}{d_e} 
 \langle \Phi_0 | H_{EDM} | \Phi_0 \rangle \nonumber \\
  &=& \frac{1}{d_e} \sum_j \langle \phi_j | h_{EDM} | \phi_j \rangle \nonumber \\
    &=& \frac{1}{d_e} \langle  \phi_v| h_{EDM} | \phi_v \rangle \nonumber \\
 &=& \frac{1}{d_e} \sum_k \sum_l C_k C_l \langle \chi_{v,k} | h_{EDM} | \chi_{v,l} \rangle .
\label{eeffdf}
\end{eqnarray}
In above expression, the sum over all MO contributions boils down to only the valence molecular orbital (SOMO, denoted as $v$ in the above set of equations) term due to the fact that contributions from the orbitals with opposite spin components of the closed-shell configuration cancel out each other. In the last line, we have expanded the valence molecular orbital as the sum of atomic orbitals (AOs), $| \chi_{v,i} \rangle$, where $i$ can be $l$ or $k$. Further details about this decomposition of SOMO to AOs can be found in, for example, Ref.~\cite{HgXanalysis}. Note that the SOMO contains contributions from both the constituent atoms of a molecule. Of all the terms in Eq.~(\ref{eeffdf}), the contributions from the $s$ and $p_{1/2}$ orbitals of Lr dominates, and accounts for about 232, 214, and 281 GV/cm, for LrO, LrF$^+$, and LrH$^+$, respectively, as shown in Fig.~\ref{fig1}. Note that the DF values for these systems are 235, 215, and 284 GV/cm for LrO, LrF$^+$, and LrH$^+$, respectively. The other contributions, such as those from $p_{3/2}$ and $d_{3/2}$ of Lr, $s$ and $p_{1/2}$ of the lighter atom etc, are less than 1 GV/cm. It is also worth noting that while the DF values of LrO and LrH$^+$ themselves are different only by about 50 GV/cm, the total effective electric fields are apart by over 80 GV/cm. This is attributed to the significant cancellation between the $OT_1$ + h.c. and the $T_1^\dag O T_1$ terms in LrO, as shown in Table \ref{termbyterm}.  

Using the values of PDM and bond-length of a molecule, we estimate the polarizing electric field for that system, which is given by  $E_{pol}=\frac{2B}{\mu}$, with $B$ being the rotational constant. The $E_{pol}$ of LrO is 18.38 kV/cm, while they are 5.79 kV/cm and 101.03 kV/cm in LrF$^+$ and LrH$^{+}$, respectively. The $E_{pol}$ required to polarize LrO and LrF$^+$ are practically achievable in the laboratory. The larger, and thus less desirable, value of $E_{pol}$ in LrH$^+$ can be attributed to its smaller reduced mass. Therefore, LrH$^+$ may not be as suitable as the other two considered candidates, but it can also be considered in an experiment if any alternative suitable technique to measure EDM in this ion can be found. 

We now turn our attention to estimating the production rates of Lr molecules for an EDM experiment. We propose to use the RIKEN heavy-ion linear accelerator (RILAC) facility because a high intensity ion beam is readily available to produce atoms of interest. As for $^{256}$Lr production, we propose the $^{209}$Bi($^{48}$Ca, $1n$) reaction as a possible candidate, where a production cross-section of 60 nanobarns (nb) has been reported~\cite{Antalic}, although it is necessary to have a prospect of stable supply of $^{48}$Ca. As explained in Ref.~\cite{Sato3}, the $^{249}$Cf($^{11}$B, $4n$) reaction was employed in the single ion beam production of $^{256}$Lr at JAEA due to its relatively high cross-section of 122 nb. The $^{249}$Cf target material is, however, radioactive, and too rare to prepare a sufficiently large target, which can be applied to a beam from the RILAC. On the other hand, $^{209}$Bi is stable and easy to handle to make a target with a large area. A typical target thickness is 300 $\mu g cm^{-2}$~\cite{Kaji}. The RILAC facility can typically provide a $3-p \mu A$ $^{48}$Ca beam. Under the situation, Lr atoms can be produced with a rate of one atom per second. In the case that GARIS (GAs-filled Recoil Ion Separator) is applied to mass-separation and single ion beam production, a transparent efficiency of 50\% is expected \cite{Morita}. In addition, in preliminary experiments at JAEA, about 20\% of Lr can be converted to LrO. Thus, we estimate N $\approx$ 0.1 molecules per second. It means only about one-atom-per-minute molecular beam could be produced, which presents major challenges for experiments with neutral molecules. On the other hand, molecular ions such as LrF$^+$ and LrH$^+$ can be efficiently guided and trapped by electromagnetic fields, enabling experiments even with just a single molecular ion.

\section{Conclusion}

We have scrutinized the sensitivity of LrF$^+$, LrH$^+$, and LrO to symmetry-violating properties. Firstly, we showed that bound states with a stable minimum does occur in the chosen molecular systems. Thereafter, we reported the computed values of $\mathcal{E}_{\mathrm{eff}}$, $\mu$, and $W_s$ for the aforementioned molecules. We observed that the values of $\mathcal{E}_{\mathrm{eff}}$ for the three superheavy molecules are about 3-4 times larger than other molecules on which EDM experiments are being performed. Similarly, the $W_s$ values were found to be about 5 times larger. We discuss a feasible pathway to produce Lr atoms and hence produce Lr molecules. We also study the properties of LrF$^+$ and LrH$^+$ molecules and their potential for future single ion experiments. Our precisely estimated bond lengths and $\mu$ values of the LrF$^+$, LrH$^+$, and LrO molecules can also be useful to guide other experimental setup using these superheavy molecules.

\section*{Acknowledgement}
Most of the calculations reported in this work were performed using the Vikram-100 high-performance computing facility at PRL, Ahmedabad. The rest of the calculations were carried out on National Supercomputing Mission's (NSM) computing resource, `PARAM Siddhi-AI', at C-DAC Pune, which is implemented by C-DAC and supported by the Ministry of Electronics and 
Information Technology (MeitY) and Department of Science and Technology (DST), Government of India. This work was partially supported by the  Office  of  Nuclear Physics,  U.S. Department of Energy,  under grants DE-SC0021176 and DE-SC0021179.

\end{document}